\pdfoutput=1

\documentclass[10pt, conference, compsocconf]{IEEEtran}
\usepackage[pdftex]{graphicx}
\ifCLASSINFOpdf
\else
\fi
\usepackage[cmex10]{amsmath}
\usepackage{url}

\def\Mpch{~h^{-1} {\rm Mpc}}
\newcommand{\cgal}{\texttt{CGAL}\ }

\begin{document}
%
\title{Alpha Shape Topology of the Cosmic Web}


\author{\IEEEauthorblockN{Rien van de Weygaert, Erwin Platen}
\IEEEauthorblockA{Kapteyn Astronomical Institute\\
University of Groningen\\
P.O. Box 800, 9700 AV Groningen, the Netherlands\\
Email: weygaert@astro.rug.nl}
\and
\IEEEauthorblockN{Gert Vegter, Bob Eldering, Nico Kruithof}
\IEEEauthorblockA{Johann Bernoulli Institute for Mathematics and Computer Science\\
University of Groningen\\
P.O. Box 407, 9700 AK Groningen, the Netherlands\\
Email: G.Vegter@gmail.com}
}

%


\maketitle

\begin{abstract}
We study the topology of the Megaparsec Cosmic Web on the basis of the Alpha Shapes 
of the galaxy distribution. The simplicial complexes of the alpha shapes are used 
to determine the set of Betti numbers ($\beta_{\rm k},k=1,\ldots,D$), which 
represent a complete characterization of the topology of a manifold. This forms 
a useful extension of the geometry and topology of the galaxy distribution by 
Minkowski functionals, of which three specify the geometrical structure of 
surfaces and one, the Euler characteristic, represents a key aspect of 
its topology. In order to develop an intuitive understanding for the relation 
between Betti numbers and the running $\alpha$ parameter of the alpha shapes, 
and thus in how far they may discriminate between different topologies, 
we study them within the context of simple heuristic Voronoi clustering models. 
These may be tuned to consist of a few or even only one specific morphological 
element of the Cosmic Web, ie. clusters, filaments or sheets. 
\end{abstract}

\begin{IEEEkeywords}
Cosmology: theory - large-scale structure of Universe - Methods: data analysis - 
techniques: image processing - Computational Geometry: tessellations - Computational 
Topology
\end{IEEEkeywords}

%
\IEEEpeerreviewmaketitle

\section{Introduction: the Cosmic Web}
\noindent The large scale distribution of matter revealed by galaxy surveys features a complex 
network of interconnected filamentary galaxy associations. This network, which has become known as 
the {\it Cosmic Web} \cite{bondweb1996}, contains structures from a few megaparsecs\footnote{The main 
measure of length in astronomy is the parsec. Technically a parsec is the distance at which we would 
see the distance Earth-Sun at an angle of 1 arcsec. It is equal to 3.262 lightyears 
$=3.086\times 10^{13} \hbox{\rm km}$. Cosmological distances are substantially larger, so that a 
Megaparsec ($=10^6\,pc$) is the regular unit of distance. Usually this goes along with $h$, the 
cosmic expansion rate (Hubble parameter) $H$ in units of $100$ km/s/Mpc ($h\approx 0.71$).} up to tens 
and even hundreds of Megaparsecs of size. Galaxies and mass exist in a wispy weblike spatial arrangement 
consisting of dense compact clusters, elongated filaments, and sheetlike walls, amidst large near-empty 
void regions, with similar patterns existing at earlier epochs, albeit over smaller scales. The 
hierarchical nature of this mass distribution, marked by substructure over a wide range of scales and 
densities, has been clearly demonstrated \cite{weybond2008b}. Its appearance has been most dramatically 
illustrated by the recently produced maps of the nearby cosmos, the 2dFGRS, the SDSS and the 2MASS redshift 
surveys \cite{colless2003,tegmark2004,huchra2005}\footnote{Because of the expansion of the Universe, 
any observed cosmic object will have its light shifted redward: its redshift $z$. According to Hubble's 
law, the redshift $z$ is directly proportional to the distance $r$ of the object, for $z\ll 1$: $cz=Hr$ 
(with $c$ the velocity of light, and $H\approx 71 km/s/Mpc$ the Hubble constant). Because it is extremely 
cumbersome to measure distances $r$ directly, cosmologists resort to the expansion of the Universe and 
use $z$ as a distance measure. Because of the vast 
distances in the Universe, and the finite velocity of light, the redshift $z$ of an object may also be 
seen as a measure of the time at which it emitted the observed radiation. }. 

The vast Megaparsec cosmic web is one of the most striking examples of complex geometric patterns 
found in nature, and certainly the largest in terms of sheer size. Computer simulations suggest that 
the observed cellular patterns are a prominent and natural aspect of cosmic structure formation through 
gravitational instability \cite{peebles80}, the standard paradigm for the emergence of structure in 
our Universe \cite{weybond2008a,springmillen2005}. 


\begin{figure*}[t]
\begin{center}
\vskip -1.0truecm
\includegraphics[bb=0 0 541 828,height=16.0cm]{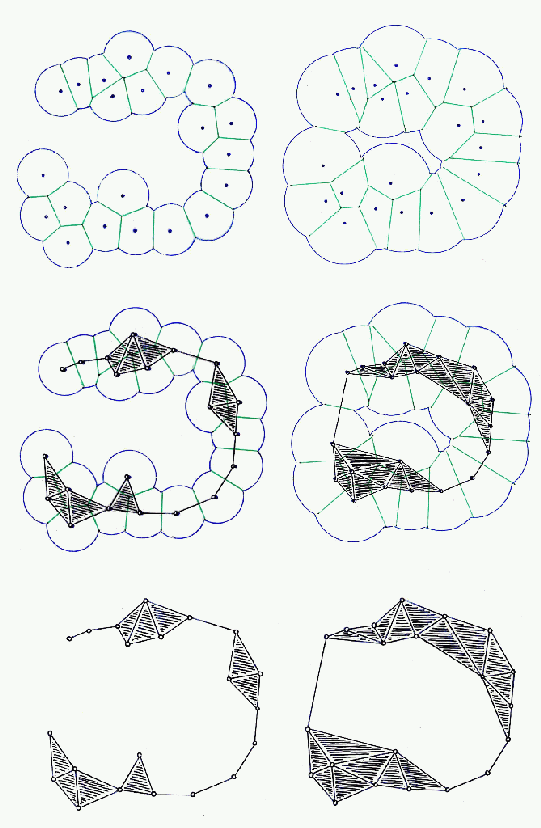} 
\vskip -0.5truecm
\caption{Illustration of Alpha Shapes. For two different values of $\alpha$, the figure shows the relation between 
the 2-D point distribution, the value of $\alpha$ and the resulting alpha shape. Around each point in the 
point sample, circles of radius $R^2=\alpha$ are drawn. The outline of the corresponding Voronoi tessellation 
within the space covered by the circles is indicated by the edges (top). All Delaunay simplices - vertices, 
edges and cells - entirely located within this space are shown in black (centre). The final resulting alpha shape 
is shown in the bottom panel. Left: small $\alpha$ value. Right: large $\alpha$ value. Based on transparencies of 
H. Edelsbrunner, used at the Jigsaw conference, Leiden, 2006.}
\label{fig:alphashape}
\end{center}
\end{figure*}

\subsection{Web Analysis}
Despite the multitude of elaborate qualitative descriptions it has remained a major challenge to characterize the 
structure, geometry and topology of the Cosmic Web. Many attempts to describe, let alone identify, the features and 
components of the Cosmic Web have been of a rather heuristic nature. The overwhelming complexity of both the individual 
structures as well as their connectivity, the lack of structural symmetries, its intrinsic multiscale nature and the wide range 
of densities that one finds in the cosmic matter distribution has prevented the use of simple and straightforward 
instruments. 

In the observational reality galaxies are the main tracers of the cosmic web and it is mainly through 
the measurement of the redshift distribution of galaxies that we have been able to map its structure. Likewise, 
simulations of the evolving cosmic matter distribution are almost exclusively based upon N-body particle computer 
calculations, involving a discrete representation of the features we seek to study. Both the galaxy distribution 
as well as the particles in an N-body simulation are examples of {\it spatial point processes} in that they 
are {\it discretely sampled} and have an {\it irregular spatial distribution}.

\medskip
\noindent {\it Topology of the Cosmic Density Field}\\
For furthering our understanding of the Cosmic Web, and to investigate its structure and dynamics, it is of prime importance 
to have access to a set of proper and objective analysis tools. In this contribution we address the topological and 
morphological analysis of the large scale galaxy distribution. To this end, we focus in particular on the alpha shapes 
of the galaxy distribution, one of the principal concepts from the field of Computational Topology \cite{edelsbrunner1983}.

In numerous previous studies, the topology and geometry of the cosmic matter distribution have been adressed in a variety of ways. A direct probe of 
the shape of the local matter distribution is the statistical distribution of inertial moments \cite{babul1992,vishniac1995,basilakos2001}. 
These concepts are closely related to the full characterization of the local geometry of the matter distribution in terms of four 
Minkowski functionals \cite{mecke1994,schmalzing1999}. These are the volume, surface area, integrated mean curvature and Euler characteristic 
of the enclosed density surfaces. The Minkowski functionals are solidly based in the theory of spatial statistics and also have 
the great advantage of being known analytically in the case of Gaussian random fields. In particular, the Euler characteristic or the 
closely related \textit{genus} of the density field has received substantial attention as a strongly discriminating factor between 
intrinsically different spatial patterns \cite{gott1986,hoyle2002}. 

The Minkowski functionals provide global characterisations of structure. An attempt to extend its scope towards providing 
locally defined topological measures of the density field has been developed in the SURFGEN project defined by Sahni and Shandarin and 
their coworkers \cite{sahni1998,shandarin2004}. It involves a local topology characterization in terms of Shapefinders, the ratios 
of Minkowski functionals. The main problem remains the user-defined, and thus potentially biased, nature of 
the continuous density field inferred from the sample of discrete objects. The usual filtering techniques suppress substructure on a 
scale smaller than the filter radius, introduce artificial topological features in sparsely sampled regions and diminish the flattened 
or elongated morphology of the spatial patterns. Quite possibly the introduction of more advanced geometry based methods to trace 
the density field may prove a major advance towards solving this problem (see contribution by Arag\'on-Calvo \& Shandarin in this volume). 
Martinez \cite{martinez2005} and Saar \cite{saar2007} have generalized the use of Minkowski Functionals by calculating their 
values in a hierarchy of scales generated from wavelet-smoothed volume limited subsamples of the 2dF catalogue. This approach 
is particularly effective in dealing with non-Gaussian point distributions since the smoothing is not predicated on the use of 
Gaussian smoothing kernels.

\medskip
\noindent {\it Alpha Shapes}\\
While most of the above topological techniques depend on some sort of user-specific smoothing and related threshold 
to specify surfaces of which the topology may be determined. An alternative philosophy is to try to let the 
point or galaxy distribution define its own natural surfaces. This is precisely where {\it Alpha Shapes} enter 
the stage. 

Alpha Shapes may be invoked to compute a large variety of geometrical and topological measures of a point distribution. 
Here we focus in particular on the determination of the Betti numbers to characterize the topology of the 
weblike spatial patterns sampled by a discrete point - i.e. galaxy - distribution. In essence, the Betti numbers 
$\beta_k$ counts the number of k-dimensional holes in an alpha complex. The first three Betti numbers, whose behaviour 
we will assess here, specify the number of individual complex ($\beta_0$), the number of independent tunnels ($\beta_1$) 
and the number of enclosed voids ($\beta_2$). Betti numbers contain more detailed topological information than Minkowski 
functionals, but unlike the latter are not sensitive to the actual geometry of an alpha shape. In order to build up an 
intuitive understanding of the behaviour of Betti numbers with respect to the weblike configurations encountered 
in the Megaparsec Universe, or in simulations of its formation, we here present a study of simpler heuristic 
Voronoi clustering models. On the basis of their topological simplicity we seek to connect the behaviour of Betti numbers 
to distinct morphological elements of the large scale Universe, such as filaments, walls, clusters and voids. 

\section{Alpha Shapes}
\label{sec:alphashape}
\noindent {\it Alpha Shape} is a description of the (intuitive) notion of the shape of a discrete point set. 
{\it Alpha Shapes} of a discrete point distribution are subsets of a Delaunay triangulation and were introduced 
by Edelsbrunner and collaborators \cite{edelsbrunner1983,mueckephd1993,edelsbrunner1994,edelsbrunner2002} 
(for a recent review see \cite{edelsbrunner2009}, and the excellent book by Edelsbrunner \& Harer 2010 \cite{edelsbrunner2010} for 
a thorough introduction to the subject). Alpha Shapes are generalizations of the convex hull of a point set and are concrete 
geometric objects which are uniquely 
defined for a particular point set. Reflecting the topological structure of a point distribution, it is one of the most 
essential concepts in the field of Computational Topology \cite{dey1998,vegter2004,zomorodian2005}. Connections to diverse 
areas in the sciences and engineering have developed, including the pattern recognition, digital shape 
sampling and processing and structural molecular biology \cite{edelsbrunner2009}. 

Applications of alpha shapes have as yet focussed on biological systems. Their main application has been in characterizing the 
topology and structure of macromolecules. The work by Liang and collaborators \cite{edelsbrunner1998,liang1998a,liang1998b,
liang1998c} uses alpha shapes and betti numbers to assess the voids and pockets in an effort to classify complex protein structures, 
a highly challenging task given the 10,000-30,000 protein families involving 1,000-4,000 complicated folds. Given the interest 
in the topology of the cosmic mass distribution \cite{gott1986,mecke1994,schmalzing1999}, it is evident that {\it alpha shapes} 
also provide a highly interesting tool for studying the topology of the galaxy distribution and N-body simulations of cosmic 
structure formation. Directly connected to the topology of the point distribution itself it would discard the need of 
user-defined filter kernels. 

\subsection{Alpha Complex and Alpha Shape: definitions}
Figure~\ref{fig:alphashape} provides a direct impression and illustration of the concept of alpha shapes, 
based on hand-drawn slides by Edelsbrunner\footnote{We kindly acknowledge permission by Herbert Edelsbrunner for the  
use of these drawings.}. If we have a point set $S$ and its corresponding Delaunay triangulation, we may identify all 
{\it Delaunay simplices} -- tetrahedra, triangles, edges, vertices -- of the triangulation. For a given non-negative 
value of $\alpha$, the {\it Alpha Complex} of a point set consists of all simplices in the Delaunay triangulation 
which have an empty circumsphere with squared radius less than or equal to $\alpha$, 
\begin{equation}
R^2\,\leq\,\alpha\,. 
\end{equation}
\noindent Here ``empty'' means that the open sphere does not include any points of $S$. For an extreme value $\alpha=0$ the alpha 
complex merely consists of the vertices of the point set. The set also defines a maximum value $\alpha_{\rm max}$, such that 
for $\alpha \geq \alpha_{\rm max}$ the alpha shape is the convex hull of the point set. 

The {\it Alpha Shape} is the union of all simplices of the alpha complex. Note that it implies that although the alpha shape 
is defined for all $0\leq \alpha < \infty$ there are only a finite number of different alpha shapes for any one point set. 
The alpha shape is a polytope in a fairly general sense, it can be concave and even disconnected. Its components can be 
three-dimensional patches of tetrahedra, two-dimensional ones of triangles, one-dimensional strings of edges and 
even single points. The set of all real numbers $\alpha$ leads to a family of shapes capturing the intuitive notion of 
the overall versus fine shape of a point set. Starting from the convex hull of a point set and gradually decreasing $\alpha$ 
the shape of the point set gradually shrinks and starts to develop cavities. These cavities may join to form tunnels and 
voids. For sufficiently small $\alpha$ the alpha shape is empty. 

The process of defining, for two different values of $\alpha$, the alpha shape for a 2-dimensional point sample is 
elucidated in fig.~\ref{fig:alphashape}. Note that the alpha shape process is never continuous: it proceeds discretely 
with increasing $\alpha$, marked by the addition of new Delaunay simplices once $\alpha$ exceeds the corresponding 
level. 

\begin{figure*}[t]
\begin{center}
  \vskip -0.5truecm
  \includegraphics[bb=0 0 500 189,width=18.0truecm]{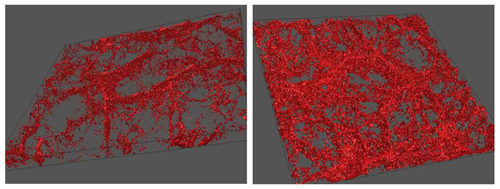} 
\vskip 0.25truecm
\caption{Examples of {\it alpha shapes} of the LCDM GIF simulation. Shown are central slices through 
two alpha shapes (top: low alpha; bottom: high alpha). The image shows the sensitivity of alpha shapes 
to the topology of the matter distribution. From: Vegter et al. 2010.}
\label{fig:gifalphashape}
\end{center}
\end{figure*}

\subsection{Alpha Shape and Topology: Betti numbers}
Following the description above, one may find that alpha shapes are intimately related to the topology of a point set. 
As a result they form a direct and unique way of characterizing the topology of a point distribution. A complete 
quantitative description of the topology is that in terms of Betti numbers $\beta_{\rm k}$ and these may indeed 
be directly inferred from the alpha shape. 

The Betti number $\beta_{\rm p}$ can be considered as the number of p-dimensional holes of an object or space. 
Formally, they are the rank of the {\it homology groups} $H_p$. There is one homology group $H_p$ per dimension $p$, 
and its rank is the {\it p-th Betti number} $\beta_{\rm p}$. The first Betti number $\beta_0$ specifies the number of 
independent components of an object. In second Betti number, $\beta_1$, may be interpreted as the number of independent 
tunnels, and $\beta_2$ as the number of independent enclosed voids. Tunnels are formed when at a certain $\alpha$ value an 
edge is added between two vertices that were already connected. When new faces are added, a tunnel can be filled 
and destroyed and thus leads to the decrease of $\beta_1$. Holes are completely surrounded by a surface or faces and 
disappears when cells are added to the alpha shape. 

The Betti numbers completely specify the topology of a manifold in terms of its connectivity. In this sense, they 
extend the principal topological characterization known in a cosmological context. Numerous cosmological studies 
have considered the {\it genus} of the isodensity surfaces defined by the Megaparsec galaxy distribution 
\cite{gott1986,hoyle2002}. The genus $g$ specifies the number of handles defining a surface and has a 
direct and simple relation to the Euler characteristic $\chi$ of the manifold, one of the Minkowski 
functionals. For a manifold consisting of $c$ components we have
\begin{equation}
g\,=\,c\,-\,{1 \over 2}\,\chi\,,
\end{equation}
where $\chi$ is the integrated intrinsic curvature of the surface, 
\begin{equation}
\chi\,=\,{\displaystyle 1 \over \displaystyle 2\pi}\,\oint\,\left({\displaystyle 1 \over \displaystyle R_1 R_2}\right)\,dS\,. 
\end{equation}
Indeed, it is straightforward to see that the topological information contained in the Euler characteristic is 
also represented by the Betti numbers, via an alternating sum relationship. For three-dimensional space, this 
is
\begin{equation}
\chi\,=\,2\,(\beta_0-\beta_1+\beta_2) 
\end{equation}
\noindent While the Euler characteristic and the Betti numbers give information about the connectivity of a manifold, 
the other three Minkowski functionals are sensitive to local manifold deformations. The Minkowski functionals therefore 
give information about the geometric and topological properties of a manifold, while the Betti numbers focus only 
on its topological properties. However, while the Euler characteristic only ``summarizes'' the topology, the 
Betti numbers represent a full and detailed characterization of the topology: homeomorphic surfaces will have the 
same Euler characteristic but two surfaces with the same $\chi$ may not in general be homeomorphic !!!

\subsection{Computing Betti numbers}
For simplicial complexes like Delaunay tessellations and Alpha Shapes, the Betti numbers can be defined on the basis 
of the oriented $k$-simplices. For such simplicial complexes, the Betti numbers can be computed by counting the number 
of $k$-cycles it contains. For a three-dimensional alpha shape, a three-dimensional simplicial complex, we can 
calculate the Betti numbers by cycling over all its constituent simplices. To this end, we base the calculation on the 
following considerations. When a vertex is added to the alpha complex, a new component is created and $\beta_0$ increases by 1. 
Similarly, if an edge is added, $\beta_1$ is increased by 1 if it creates a new cycle, which would be an increase in the 
number of tunnels. Otherwise, two components get connected so that the number of components is decreased by one: 
$\beta_0$ is decreased by 1. If a face is added, the number of holes is increased by one if it creates a new cycle. 
Otherwise, a tunnel is filled, so that $\beta_1$ is decreased by one. Finally, when a (tetrahedral) cell is added, 
a hole is filled up and $\beta_2$ is lowered by 1. 

Following this procedure, the algorithm has to include a technique for determining whether a $k$-simplex belongs to a
$k$-cycle. For vertices and cells, and thus 0-cycles and 3-cycles, this is rather trivial. For the detection of 
1-cycles and 2-cycles we used a somewhat more elaborate procedure involving union-finding structures \cite{delfinado1993}. 

\subsection{Computational Considerations}
\noindent For the calculation of the alpha shapes of the point set we resort to the Computational Geometry 
Algorithms Library, \cgal\footnote{\cgal is a \texttt{C++} library of algorithms and
data structures for Computational Geometry, see \url{www.cgal.org}.}. Within this context, Caroli \& Teillaud 
recently developed an efficient code for the calculation of two-dimensional and three-dimensional alpha shapes 
in periodic spaces.

The routines to compute the Betti numbers from the alpha shapes were developed within our own project. 

\section{Alpha Shapes of the Cosmic Web}
\noindent In a recent study, Vegter et al. computed the alpha shapes for a set of GIF simulations 
of cosmic structure formation \cite{eldering2006,vegter2010}. It concerns a $256^3$ particles GIF $N$-body simulation, 
encompassing a $\Lambda$CDM ($\Omega_m=0.3,\Omega_{\Lambda}=0.7,H_0=70\, {\rm km/s/Mpc}$) density field within a 
(periodic) cubic box with length $141h^{-1} {\rm Mpc}$ and produced by means of an adaptive 
${\rm P^3M}$ $N$-body code \cite{kauffmann1999}. 

Fig.~\ref{fig:gifalphashape} illustrates the alpha shapes for two different values of $\alpha$, 
for two-dimensional section through the GIF simulation. The top panel concerns a low value of 
$\alpha$, the bottom one a high value. The intricacy of the weblike patterns is very nicely 
followed. The low alpha configuration highlights the interior of filamentary and 
sheetlike features, and reveals the interconnection between these major structural elements. 
The high value alpha shape not only covers an evidently larger volume, but does so by connecting 
to a lot of finer features in the Cosmic Web. Noteworthy are the tenuous filamentary and 
planar extensions into the interior of the voids.  

These images testify of the potential power of alpha shapes in analyzing the weblike cosmic 
matter distribution, in identifying its morphological elements, their connections and in particular 
also its hierarchical character. 

However, to understand and properly interpret the topological information contained in these 
images we need first to assess their behaviour in simpler yet similar circumstances. To this 
end, we introduce a set of heuristic spatial matter distributions, Voronoi clustering models.

\begin{figure*}[t]
  \centering
     \vskip -0.5truecm
     \includegraphics[bb=0 0 539 182,width=16.8truecm]{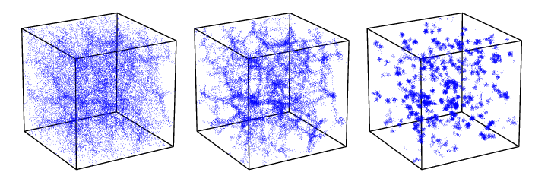} 
     \vskip -0.0truecm
     \caption{\small Three different patterns of Voronoi element galaxy distributions, 
           shown in a 3-D cubic setting. The depicted spatial distributions 
           correspond to a wall-dominated Voronoi Universe (left), a filamentary 
           Voronoi Universe (centre) and a cluster-dominated Voronoi Universe 
           (right). }  
\label{fig:vorelm3}
\end{figure*}

\section{Voronoi Clustering Models}
\noindent {\it Voronoi Clustering Models} are a class of heuristic models for cellular distributions 
of matter which use the Voronoi tessellation as the skeleton of the cosmic matter distribution 
\cite{weyicke1989,weygaert1991,weygaert2007}. 

The aspect which is modelled in great detail by Voronoi tessellations is that of the large scale 
clustering of the morphological elements of the Cosmic Web. It is the stochastic yet non-Poissonian 
geometrical distribution of the {\it walls}, {\it filaments} and {\it clusters} which generates 
the large-scale clustering properties of matter and the related galaxy populations. 

The small-scale distribution of galaxies, i.e. the distribution within the various 
components of the cosmic skeleton, involves the complicated details of highly 
nonlinear small-scale interactions of the gravitating matter. Well-defined 
and elaborate physical models and/or N-body computer simulations might fill in this 
aspect, although it would lead away from the Voronoi model's true purpose and conceptual 
simplicity. In the Voronoi models described here we complement the geometrically 
fixed configuration of the Voronoi tessellations with a heuristic prescription for 
the location of particles or model galaxies within the tessellation. 

\subsection{Voronoi components}
According to the Voronoi clustering models, each of the geometric elements of the 
3-D Voronoi tessellations is identified with a morphological component of the 
Cosmic Web. In table~\ref{tab:vorcomp} we have listed the various identifications.

\begin{table}[h]
\begin{center}
\begin{tabular}{||l||l||}
\hline 
\hline
&\\
\hskip 0.25truecm Geometric Component \hskip 2.0truecm & \hskip 0.5truecm Cosmic Structure \hskip 1.0truecm \\
&\\
\hline 
&\\
\hskip 0.25truecm Voronoi Cell \hskip 2.0truecm & \hskip 0.25truecm Voids, Field \hskip 1.0truecm \\
&\\
\hskip 0.25truecm Voronoi Wall \hskip 2.0truecm & \hskip 0.25truecm Walls, Sheets, Superclusters \hskip 1.0truecm \\
&\\
\hskip 0.25truecm Voronoi Edge \hskip 2.0truecm & \hskip 0.25truecm Filaments, Superclusters \hskip 1.0truecm \\
&\\
\hskip 0.25truecm Voronoi Vertex \hskip 2.0truecm & \hskip 0.25truecm Clusters \hskip 1.0truecm \\
&\\
\hline
\hline
\end{tabular}
\end{center}
\caption{Identification of geometric components in Voronoi tessellations with morphological components 
of the Cosmic Web.}
\label{tab:vorcomp}
\end{table}

\subsection{Voronoi Element and Voronoi Evolution Models}
\noindent We distinguish two different yet complementary approaches, {\it ``Voronoi Element models''} and 
{\it ``Voronoi Evolution models''}. Both the Voronoi Element Models and the Voronoi Evolution Models are obtained 
by projecting an initially random distribution of $N$ sample points/galaxies onto the walls, edges or vertices of 
the Voronoi tessellation defined by $M$ nuclei. The Voronoi Element Models do this by a heuristic and user-specified 
mixture of projections on the various geometric elements of the Voronoi skeleton. The Voronoi Evolution Models 
accomplish this via a gradual motion of the galaxies from their initial random location in their Voronoi cell, 
directed radially away from the cell's nucleus. 

\subsection{Voronoi Element Models}
\label{sec:vorelm}
\noindent {\it ``Voronoi Element models''} are fully heuristic models. They are user-specified spatial 
galaxy distribution within the {\it cells} (field), {\it walls}, {\it edges} and {\it vertices} of 
a Voronoi tessellation. The initially randomly distributed $N$ model galaxies are projected onto the relevant 
Voronoi wall, Voronoi edge or Voronoi vertex or retained within the interior of the Voronoi cell in which 
they are located. The field galaxies define a sample of randomly distributed points throughout the entire 
model volume. The Voronoi Element Models are particularly apt for studying systematic properties of spatial 
galaxy distributions confined to one or more structural elements of nontrivial geometric spatial patterns. 

{\it Simple Voronoi Element Models} place  their model galaxies exclusively in either walls, edges or vertices. 
The versatility of the Voronoi element model also allows combinations in which field (cell), wall, filament and 
vertex distributions are superimposed. These complete composite particle distributions, {\it Mixed Voronoi 
Element Models}, include particles located 
in four distinct structural components:
\begin{enumerate}
\item[$\bullet$] \emph{Field}: \\ \ \ \ \ \ \ \ \ \  Particles located in the {\it interior of Voronoi cells}
\\ \ \ \ \ \ \ \ \ \ (i.e. randomly distributed across the entire model box)
\item[$\bullet$] \emph{Wall}: \\ \ \ \ \ \ \ \ \ \ \ \,Particles within and around the {\it Voronoi walls}.
\item[$\bullet$] \emph{Filament}: \\ \ \ \ \,Particles within and around the {\it Voronoi edges}.
\item[$\bullet$] \emph{Blobs}: \\ \ \ \ \ \ \ \ \ \ Particles within and around the {\it Voronoi vertices}.
\end{enumerate}
\noindent The characteristics of the patterns and spatial distribution in the composite Voronoi Element models can be 
varied and tuned according to the fractions of galaxies in in Voronoi walls, in Voronoi edges, in Voronoi vertices and 
in the field. These fractions are free parameters to be specified by the user. 

\begin{figure*}[t]
\begin{center}
     \includegraphics[bb=0 0 539 461,width=16.8truecm]{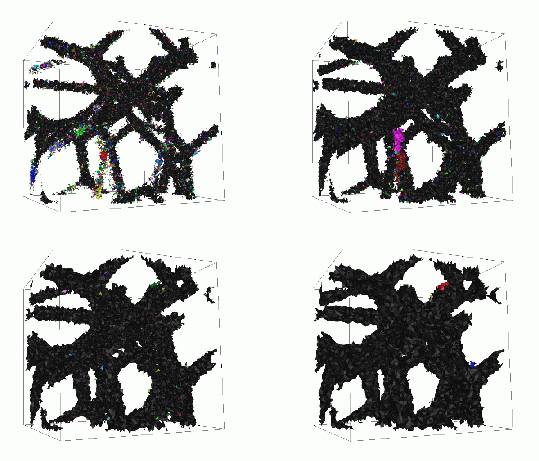} 
\vskip -0.0truecm
\caption{Four alpha shapes of a Voronoi filament model realization. It concerns a sample of 
200000 particles in a periodic box of $50\Mpch$ size with 8 Voronoi cells. From top left to 
bottom right: $\alpha$=$0.5\times 10^{-4}$, $1.0\times 10^{-4}$, $2 \times 10^{-4}$ and 
$4.0 \times 10^{-4}$. See text.}
\label{fig:vorfilalpha}
\end{center}
\end{figure*}

In fig.~\ref{fig:vorelm3} we have shown three different three-dimensional {\it Simple Voronoi Element Model} galaxy 
distributions. The lefthand model realization corresponds to the model in which galaxies are exclusively 
located inside walls, a second one where these are concentrated in and around filaments and the third one 
restricted to galaxies located within clusters. 

\begin{figure*}
\vskip -1.0truecm
\begin{center}
\includegraphics[bb=0 0 835 1024,height=0.75\textheight]{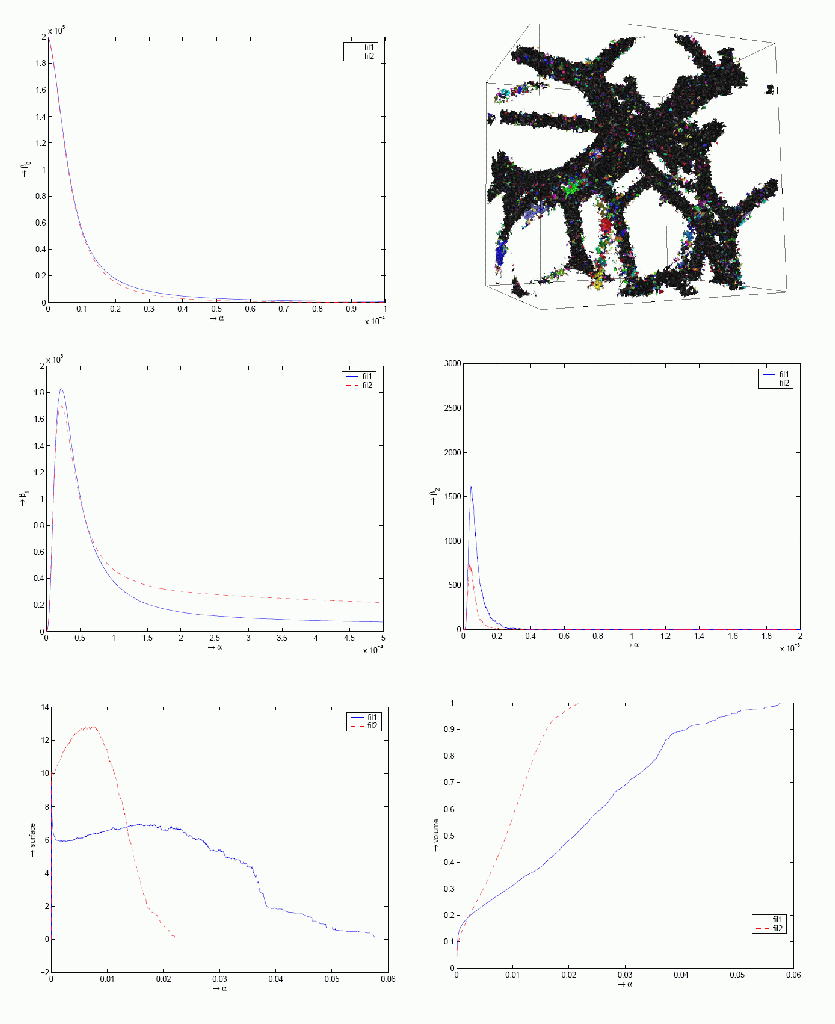} 
\vskip -0.0truecm
\caption{Topological and Geometric parameters Voronoi filament model. For a Voronoi filament model we show an 
example of an alpha shape (top right), along with the behaviour of five topological parameters as a function of 
$\alpha$. The different colours of the alpha shape represent the different connected components. 
Top left: $\beta_0$, the number of components of the alpha shape. Centre left: $\beta_1$, the number of 
tunnels in the alpha shape. Centre right: $\beta_2$, the number of (surrounded) holes in the alpha shape. 
Bottom left: total surface area of the alpha shapes. Bottom right: total volume of the alpha shapes. Blue lines:  
realization with 8 Voronoi cells in box. Red lines: realization with 64 cells in box.}
\label{fig:vorfiltop}
\end{center}
\end{figure*}

\subsection{Voronoi Evolution Models}
\label{sec:vorkinm}
The second class of Voronoi models is that of the {\it Voronoi Evolution models}. They attempt to 
provide weblike galaxy distributions that reflect the outcome of realistic cosmic structure formation 
scenarios. They are based upon the notion that voids play a key organizational role in the development 
of structure and makes the Universe resemble a soapsud of expanding bubbles \cite{icke1984}. 
While the galaxies move away from the void centres, and stream out of the voids towards the sheets, 
filaments and clusters in the Voronoi network the fraction of galaxies in the voids (cell interior), 
the sheets (cell walls), filaments (wall edges) and clusters (vertices) is continuously changing 
and evolving. The details of the model realization depends on the time evolution specified by the 
particular Voronoi Evolution Model. 

Within the class of Voronoi Evolution Models the most representative and most frequently used are the 
{\it Voronoi kinematic models}. They form the idealized and asymptotic description of the outcome of 
hierarchical gravitational structure formation process, with single-sized voids forming around depressions in 
the primordial density field. The {\it Voronoi Kinematic Model} ``simulates'' the asymptotic weblike galaxy 
distribution implied by the hierarchical void formation process by assuming a single-size dominated void population. 
It is based upon the notion that voids play a key organizational role in the development of structure and makes the 
Universe resemble a soapsud of expanding bubbles \cite{icke1984}, forming voids forming around a dip in the primordial 
density field. 

This is translated into a scheme for the displacement of initially randomly distributed galaxies within the 
Voronoi skeleton. Within a void, the mean distance between galaxies increases uniformly in the course of time. When 
a galaxy tries to enter an adjacent cell, the velocity component perpendicular to the cell wall disappears. Thereafter, 
the galaxy continues to move within the wall, until it tries to enter the next cell; it then loses its velocity
component towards that cell, so that the galaxy continues along a filament. Finally, it comes to rest in a 
node, as soon as it tries to enter a fourth neighbouring void. 

\bigskip
\noindent {\it Kinematic Model Configurations}\\
The resulting evolutionary progression within the Voronoi kinematic scheme proceeds 
from an almost featureless random distribution towards a distribution in which matter 
ultimately aggregates into conspicuous compact cluster-like clumps.

The steadily increasing contrast of the various structural features is accompanied by a 
gradual shift in topological nature of the distribution. The virtually uniform and featureless 
particle distribution at the beginning ultimately unfolds into a highly clumped distribution of almost 
only clusters (vertices). This evolution involves a gradual progression via a wall-like through a 
filamentary towards an ultimate cluster-dominated matter distribution. By then nearly all matter has 
streamed into the nodal sites of the cellular network.

\section{Topological Analysis of Voronoi Clustering Models}
Our topological analysis consists of a study of the systematic behaviour of the 
Betti numbers $\beta_0$, $\beta_1$ and $\beta_2$, and the surface and volume of the 
alpha shapes of Voronoi clustering models (see \cite{eldering2006,vegter2010}). For 
each point sample we investigate the alpha shape for the full range of $\alpha$ parameters.

We generated 12 Voronoi clustering models. Each of these contained 200000 particles 
within a periodic box of size $100 \Mpch$. Of each configuration, we made two 
realizations, one with 8 centres and Voronoi cells and one with 64. One class of 
models consisted of pure Voronoi element models. One of these 
is a pure Voronoi wall model, in which all particles are located within the 
walls of the Voronoi tessellation. The other is a pure filament model. In addition, 
there are 4 Voronoi kinematic models, ranging from a mildly evolved to a strongly 
evolved configuration. In all situations, the clusters, filaments and walls have a 
finite Gaussian width of $R_f=1.0\Mpch$. 

An impression of the alpha shape development may be obtained from the four panels 
in fig.~\ref{fig:vorfilalpha}. Different colours depict different individual 
components of the alpha shape. For the smallest value of $\alpha$, $\alpha =0.5 \times 10^{-4}$, 
we see that the Delaunay simplices contained in the alpha shape delineate accurately 
nearly all the edges/filaments in the particle distribution. As $\alpha$ increase, going 
from the topleft panel down to the bottom right one, we find that the alpha shape fills in 
the planes of the Voronoi tessellation. For even larger values of $\alpha$, the alpha shape 
includes the larger Delaunay simplices that are covering part of the interior of 
the Voronoi cells. It is a beautiful illustration of the way in which alpha shapes define, 
as it were, naturally evolving surfaces that are sensitive to every detail of the 
morphological and topological structure of the cosmic matter distribution. 

\begin{figure*}
\begin{center}
\includegraphics[bb=0 0 1280 953,width=0.98\textwidth]{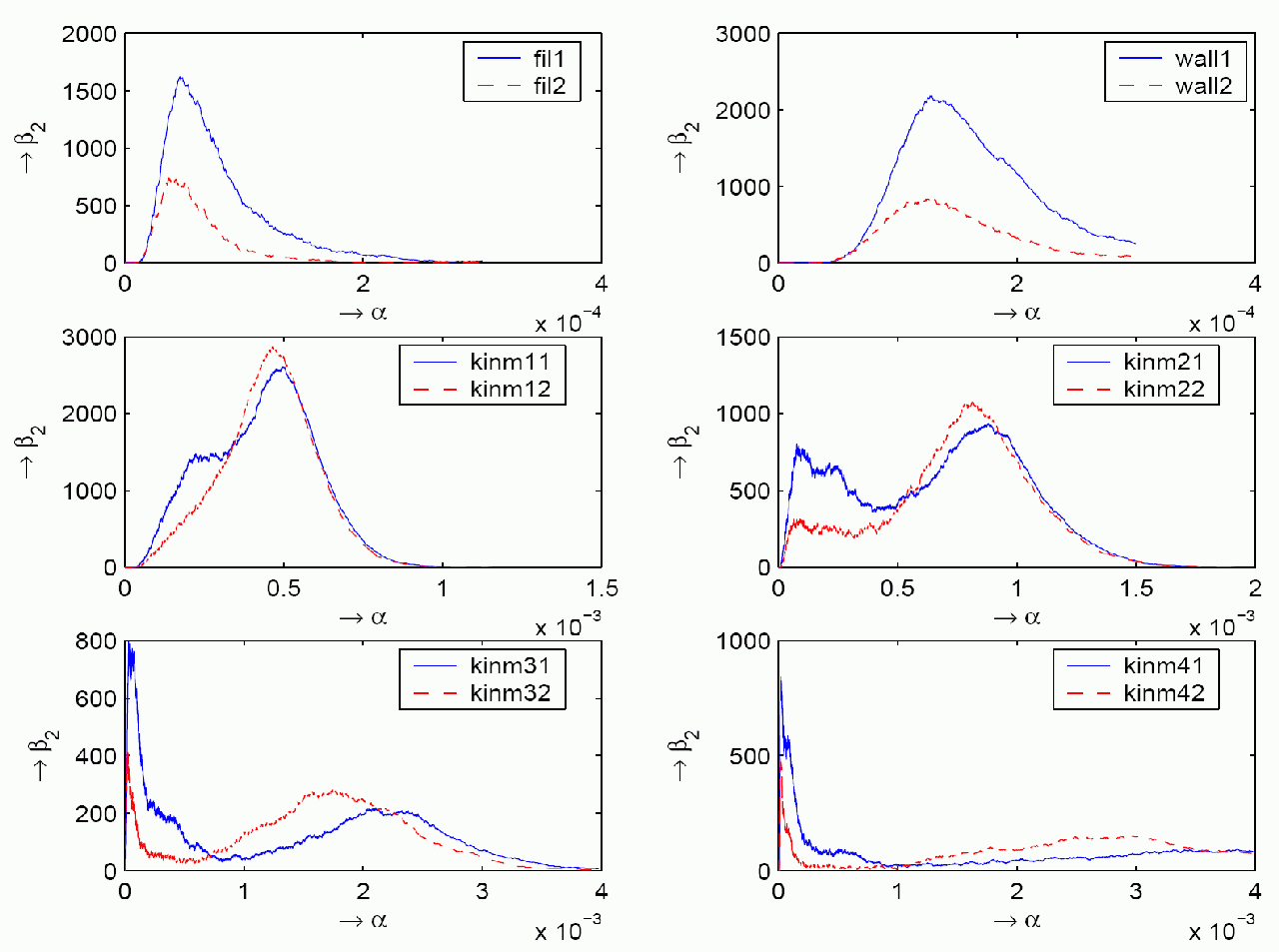} 
\vskip -0.0truecm
\caption{Third Betti number $\beta_2$ for different clustering models. Shown are the 
curves $\beta_2(\alpha)$ for six Voronoi clustering models. Top left: Voronoi filament model. 
Top right: Voronoi wall model. Centre left to Bottom right panel: four stages of the Voronoi 
kinematic model, going from a moderately evolved model dominated by walls (centre left) to a 
highly evolved one dominated by filaments and clusters (bottom right).}
\label{fig:vorclustbeta2}
\end{center}
\end{figure*}

\begin{figure*}
\vskip -0.5truecm
\begin{center}
\includegraphics[bb=0 0 640 321,width=0.9\textwidth]{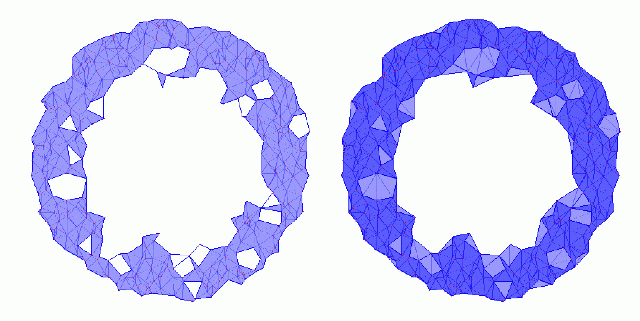} 
\vskip -0.0truecm
\caption{Cartoon illustrating the idea of persistence. Two alpha shapes corresponding to a random point 
distribution on the surface of a torus. The alpha shape on the left has 18 holes, of which only the 
central one is significant. The alpha shape on the right, corresponding to a larger $\alpha$ contains 
this one hole.}
\label{fig:persistence}
\end{center}
\end{figure*}
\subsection{Filament Model topology}
We take the Voronoi filament model as a case study. Its topology and geometry is studied 
by following the behaviour of the three Betti numbers $\beta_0$, $\beta_1$ and $\beta_2$ of 
the corresponding alpha shapes as a function of the parameter $\alpha$. Also we adress two 
of the Minkowski functionals for the alpha shapes, namely their volume and surface area 
(note that the Euler characteristic is already implicitly included in the Betti numbers). 

Fig.~\ref{fig:vorfiltop} shows the relation between the Betti numbers, surface and volume 
of the alpha shapes and the value of $\alpha$. The first Betti number, $\beta_0$, decreases 
monotonously as $\alpha$ increases. This Betti number specifies the number of isolated 
components in the alpha shape, and at $\alpha=0$ it is equal to the number of points of the 
sample, $\beta_0(\alpha=0)=200000$. At larger $\alpha$, more and more components merge 
into larger entities, which explains the gradual decrease of $\beta_0$ with increasing 
$\alpha$. 

The second Betti number, $\beta_1$, represent the number of independent tunnels. 
At first, at small values of $\alpha$ it increases steeply, as the large number of 
individual components grow as more Delaunay edges are added to them. Once 
isolated components start to fill up and merge with others, at values 
$\alpha > 0.25\times 10^{-4}$, we see a steep decrease of the Betti number $\beta_1$. 

In the case of the singly morphological distribution of the Voronoi filament model the 
third Betti number, $\beta_2$, has a behaviour resembling that of $\beta_1$: a peaked 
distribution around a finite value of $\alpha$. This Betti number represents the number 
of holes in the alpha shape and it is easy to understand that at small $\alpha$ values 
this number quickly increases as each of the individual alpha shape components gets 
extended with new Delaunay simplices. However, once $\alpha$ gets beyond a certain 
value this will include more and more tetrahedra. These start to fill up the holes. 
Notice that in the given example of the Voronoi filament model this occurs at 
$\alpha \approx 0.45\times 10 ^{-4}$ (also see fig.~\ref{fig:vorclustbeta2}), 
substantially beyond the peak in the $\beta_1$ distribution: on average larger 
values of $\alpha$ are needed to add complete Delaunay tetrahedra to 
the alpha shape. 

We also studied two additonal Minkowski functionals, volume and surface, to 
assess the geometrical properties of the evolving alpha shapes of the Voronoi 
filament models. The results are shown in the bottom panels of fig.~\ref{fig:vorfiltop}. 
Evidently, the volume of the alpha shape increases monotonously as $\alpha$ 
becomes larger and more and more tetrahedral cells become part of the alpha shape. The 
surface area has a somewhat less straightforward behaviour. Over a substantial range 
the surface area grows mildy as individual components of the alpha shape grow. When 
these components start to merge, and especially when holes within the components get 
filled up, the surface area shrinks rapidly. Once, the whole unit cube is filled, 
the surface area has shrunk to zero. 

\subsection{Betti Systematics}
Having assessed one particular Voronoi clustering model in detail, we may try to identify 
the differences between the different models. While this is still the subject of ongoing 
research, we find substantial differences between the models on a few particular 
aspects. Here we discuss two in more detail: 

\medskip
\noindent {\it Number of components: $\beta_0$}\\
While for all models the curve $\beta_0(\alpha)$ is a monotonously 
decreasing function of $\alpha$, the range over which $\beta_0$ differs substantially 
from unity and the rate of decrease are highly sensitive to the underlying distribution. 
In fact, the derivative $\partial \beta_0/\partial \alpha$ contains interesting features, 
like a minimum and a varying width, which are potentially interesting for discriminating 
between the underlying topologies.

\medskip
\noindent {\it Number of holes: $\beta_2$}\\
The most outstanding Betti number is $\beta_2$, i.e. the number of holes in an alpha shape. As one 
may infer from fig.~\ref{fig:vorclustbeta2}, substantial and systematic differences between 
different models can be observed. This concerns not only the values and range over which 
$\beta_2$ reaches a maximum, as for the pure filament and wall models, but even entirely different 
systematic behaviour in the case of the more elaborate and complex Voronoi kinematic models 
(see e.g. the review by Van de Weygaert 2010 \cite{weygaert2010} and \cite{weygaert2007}). 

In the case of the kinematic models we find more than one peak in the $\beta_2(\alpha)$ 
distribution, each corresponding to different morphological components of the particle 
distribution. In this respect, it is revealing to see follow the changes in $\beta_2(\alpha)$ 
as we look at different evolutionary stages. 

The four panels from centre left to bottom right in fig.~\ref{fig:vorclustbeta2} correspond to 
four different stages of evolution. The centre left one concerns a moderately evolved matter 
distribution, dominated by walls. The centre right panel corresponds to a stage at which walls and 
filaments are approximately equally prominent. The bottom left panel is a kinematic model in which 
filaments represent more than $40\%$ of the mass, while walls and the gradually more prominent 
clusters each represent around $25\%$ of the particles. The final bottom right diagram corresponds 
to a highly evolved mass distribution, with clusters and filaments representing each around 
$40\%$ of the particles. 

The different morphological patterns of the Voronoi kinematic models are strongly reflected 
in the behaviour of $\beta_2(\alpha)$. In the centre left panel we find a strong peak at 
$\alpha \approx 5 \times 10^{-4}$, with a shoulder at lower values. The peak reflects the 
holes defined by the walls in the distribution, while the shoulder finds its origin in the 
somewhat smaller holes defined by filaments: the average distance between walls is in the 
order of the Voronoi cell size, while the average filament distance is more related to the 
Voronoi wall size. The identity of the peaks becomes more clear when turning to the two peak distribution 
in the centre right panel, in which the strong peak  at $\alpha \approx 1 \times 10^{-4}$ is a direct 
manifestation of the strongly emerged filaments in the matter distribution. As the shift to filaments 
and clusters continues, we even see the rise of a third peak at much smaller values of 
$\alpha$ (bottom panels). This clearly corresponds to the holes in the high density and 
compact cluster regions. 

\section{Conclusions and Prospects: Persistence}
We have established the promise of alpha shapes for measuring the 
topology of Megaparsec galaxy distribution in the Cosmic Web by studying 
the Betti numbers and several Minkowski functionals of a set of heuristic 
Voronoi clustering models. Alpha Shape analysis has the great advantage of 
being self-consistent and natural, involving shapes and surfaces that are 
entirely determined by the point distribution, independent of any artificial 
filtering. 

The one outstanding issue we have not adressed is that of noise in the 
dataset. Discrete point distributions are necessarily beset by shotnoise. As 
a result, the alpha shapes will reflect the noise in the point distribution. 
The induced irregularities in the alpha shapes induce holes and tunnels which 
do not represent any real topological structure but nonetheless influence the 
values of the Betti numbers. 

Edelsbrunner et al. (2000) \cite{edelsbrunner2000} introduced the concept of {\it persistence} to seek to 
filter out the insignificant structures (see also \cite{edelsbrunner2002}, 
and \cite{edelsbrunner2010} for a detailed and insightful review). The basic idea is 
that holes and tunnels that remain in existence over a range of $\alpha$ values are 
significant and should be included in the Betti number calculation. The 
range is a user-defined persistence parameter $p$. The implementation of 
the concept of persistence in our topological study will be adressed in 
forthcoming work. 

\section*{Acknowledgement}
We are very grateful to Bernard Jones, Monique Teillaud, Manuel Caroli and Pratyush Pranav for useful 
discussions. We particularly wish to acknowledge Monique and Manuel for developing the elegant and 
efficient periodic boundary condition alpha shape CGAL software. We are also most grateful to 
Herbert Edelsbrunner for incisive comments on an earlier draft of this paper and his kind permission 
to use his handwritten transparencies for figure 1.



%

\end{document}